\documentstyle[pre,aps,multicol,graphicx,amssymb]{revtex}
\newcommand{\bleq}{\ifpreprintsty \else
\end{multicols}\vspace*{-3.5ex}{\tiny \noindent\begin{tabular}[t]{c|}
\parbox{0.493\hsize}{~} \\ \hline \end{tabular}} \fi}
\newcommand{\eleq}{\ifpreprintsty \else
{\tiny\hspace*{\fill}\begin{tabular}[t]{|c}\hline
\parbox{0.49\hsize}{~} \\
\end{tabular}}\vspace*{-2.5ex}\begin{multicols}{2} \fi}
\newcommand{\bcols}{\ifpreprintsty\else\begin{multicols}{2}\fi}
\newcommand{\ecols}{\ifpreprintsty\else\end{multicols}\fi}
\newcommand{\be}{\begin{eqnarray}} \newcommand{\ee}{\end{eqnarray}}

\begin{document}
\draft

\title{The Weakly Pushed Nature of ``Pulled'' Fronts with a Cutoff}

\author{Debabrata Panja and Wim van Saarloos} \maketitle
\begin{center}
{\em Instituut--Lorentz, Universiteit Leiden, Postbus 9506, 2300 RA
Leiden, The Netherlands}
\end{center}

\date{\today}

\vspace{-2mm}
\begin{abstract} 
The concept of pulled fronts with a cutoff $\epsilon$ has been
introduced to model the effects of discrete nature of the constituent
particles on the asymptotic front speed in models with continuum
variables (Pulled fronts are the fronts which propagate into an
unstable state, and have an asymptotic front speed  equal to the
linear spreading speed $v^*$ of small linear perturbations around the
unstable state). In this paper, we demonstrate that the introduction
of a cutoff actually makes such  pulled fronts  weakly pushed. For the
nonlinear diffusion equation with a cutoff, we show that the longest
relaxation times $\tau_m$ that govern the convergence to the
asymptotic front speed and profile, are given by $\tau_m^{-1} \simeq
[(m+1)^2-1] \pi^2 / \ln^2 \epsilon$, for $m=1,2,\cdots$.
\end{abstract} 

\maketitle

\bcols

\section{Introduction}

Pulled fronts are fronts that propagate into an unstable state, for
which the propagation dynamics is essentially that they are being
pulled along by the growth and spreading of small perturbations about
the unstable state, into which the front propagates. Concretely, this
means that their asymptotic speed $v_{as}$ is equal to the linear
spreading speed $v^*$ of perturbations around the unstable state,
$v_{as}=v^*$ \cite{dee,bj,vs2,ebert1,ebert,bd}. Fronts that propagate
into an unstable state but for which $v_{as} >v^*$ are often termed
``pushed''. The name stems from the intuitive idea
\cite{stokes,paquette} that in this regime, the dynamics in the
nonlinear front region or the bulk region behind the front  actually
{\em drives} the front propagation: effectively it pushes the front
from behind, and the front moves with a speed that is higher than the
natural speed with which small perturbations about the unstable state
spread by themselves ahead of the front.

It is clear from the definition that the concept of a pulled front
essentially pertains to a continuum formulation of the relevant
dynamical variables. The linear spreading speed $v^*$ is defined and
calculated in practice by  considering perturbations of {\em
arbitrarily small amplitude} about the unstable state of  the
dynamical equations; the value of $v^*$ then follows from an
asymptotic analysis of the linearized dynamical equations
\cite{ebert}.  However, in all cases, in which one cannot ignore the
fact that  matter is made of discrete particles,  one cannot  perturb
the unstable state by any arbitrary small amount, because this amount
must be at least one ``quantum'' of particle large.

To model this discrete nature of the constituent particles by means of
a continuum equation, Brunet and Derrida \cite{bd} studied the
nonlinear diffusion equation
\begin{eqnarray}
\frac{\partial\phi}{\partial t}\,=\,\frac{\partial^2\phi}{\partial
x^2}\,+\,f(\phi)\,,
\label{e1}
\end{eqnarray} 
with  a cutoff $\epsilon$ in the growth term  $f(\phi)$,
\begin{eqnarray}
f(\phi)\,=\,\Theta(\phi-\epsilon)\,[
\phi\,-\,\phi^n]\,,\nonumber\\&&\hspace{-2cm}\quad n>1\,,\,\,\,
\mbox{e.g.,\,\,}n\,=\,2\,\,\,\mbox{or}\,\,\,3\,.
\label{e2}
\end{eqnarray}
Without the cutoff, i.e., for $\epsilon=0$, this equation is the
well-known nonlinear diffusion equation, which has been used since
long as the simplest model to study front propagation into an unstable
state \cite{fisher,kpp,aw}. Brunet and Derrida \cite{bd} found that
the asymptotic front speed $v_{as}$ goes as
\begin{eqnarray}
v_{as}= v_\epsilon\,\simeq \,v^*\,-\,\frac{\pi^2}{\ln^2\epsilon}\,,
\label{e3}
\end{eqnarray} 
where $v^* =2$ is the asymptotic speed of the corresponding pulled
front of Eq. (\ref{e1}) for $\epsilon=0$. The above formula shows that
the front speed $v_\epsilon$ converges very slowly to the asymptotic
speed $v^*$; this illustrates that unlike pushed fronts, pulled fronts
are very sensitive to small changes in the dynamics of the phase into
which they propagate.

In comparing with stochastic models of particles on a lattice, Brunet
and Derrida  associated the cutoff $\epsilon$ with $1/N$, where $N$ is
the average number of particles in a correlation region in the
saturation phase behind the front \cite{bd}. Although the validity of
this identification has been the matter of some debate, it appears
that Eq. (\ref{e3}) with $\epsilon=1/N$ does give the proper
asymptotic correction to the front speed even for very large $N$. We
refer to the literature \cite{bd,kns,levine,vanzon,PvS} for a further
discussion of the applicability of these ideas to stochastic models.

It is intuitively clear that as soon as we introduce this cutoff,
fronts that are pulled for $\epsilon=0$, must actually become weakly
pushed as soon as $\epsilon >0$. After all, any perturbation around
the value $\phi=0$ does not start to grow until the local $\phi$ value
crosses $\epsilon$, so strictly speaking, the linear spreading
velocity $v^* (\epsilon) $ of {\em arbitrarily small}  linear
perturbations about the state $\phi=0$ vanishes. As $v_\epsilon >
v^*(\epsilon)=0$, one clearly must have a weakly pushed front. With
this idea in mind, it is natural to address the convergence of the
front speed to the asymptotic value, since it is well known that the
speed of pulled fronts relaxes algebraically slowly to the asymptotic
value $v^*$ \cite{vs2,ebert1,ebert,bd}, while pushed fronts normally
have exponential relaxation to their asymptotic speed.

These observations motivate us to investigate here the slowest
relaxation modes of the stability spectrum of fronts for the nonlinear
diffusion equation (\ref{e1}), with a cutoff $\epsilon$ in the growth
term (\ref{e2}). We calculate these relaxation modes explicitly for
small $\epsilon$, and find that the slowest relaxation times $\tau_m$
are given by
\begin{equation}
\tau^{-1}_m \simeq \frac{ \left[ (m+1)^2 -1\right] \pi^2}{\ln^2
\epsilon}  \quad\quad\quad m\,=\,1,\,2\ldots
\end{equation}
Hence, the relaxation times of the front velocity and profile approach
zero as $\epsilon \rightarrow 0$, but only logarithmically
slowly. Just like the corrections to the front speed for practical
values of $\epsilon$ are often significant, so is the exponential
relaxation, for example, for $\epsilon=10^{-5}$, the longest
relaxation time $\tau_1$ is about 4.48. Thus, while in the absence of
a cutoff the front speed is approached very slowly, only as $3/2t$
where $t$ is the time \cite{vs2,ebert1,ebert,bd}, with a realistic
value of $\epsilon$, the front speed converges relativly quickly to
the asymptotic value.

\section{Stability analysis of the asymptotic front solution}

\subsection{The Stability Operator}

The asymptotic shape of the front is a uniformly translating front
solution $\phi_\epsilon(x,t)$ which is a function of  {\it only\/} the
comoving coordinate $\xi=x-v_\epsilon t$, and which is obtained by
solving the ordinary differential equation
\begin{eqnarray}
-\,v_\epsilon\,\frac{d\phi_\epsilon(\xi)
 }{d\xi}\,=\,\frac{d^2\phi_\epsilon(\xi)}{d\xi^2}\,+\,f(\phi_\epsilon(\xi))\,.
\label{e4}
\end{eqnarray} 
In carrying out the linear stability analysis of this front solution
it is convenient to follow the standard route of transforming the
linear eigenvalue problem into a Schr\"odinger eigenvalue problem
\cite{bj,ebert}.   We consider a function $\phi(x,t)$, which is
infinitesimally different from
$\phi_\epsilon(\xi)\equiv\phi_\epsilon(x-v_\epsilon t)$ in the
comoving frame, i.e., $\phi(x,t)=\phi_\epsilon(x-v_\epsilon
t)+\eta(\xi,t)$. Upon linearizing the dynamical equation in the
comoving frame, one finds that the function
$\eta(x,t)\equiv\eta(\xi,t)$ obeys the following equation:
\begin{eqnarray}
\frac{\partial\eta}{\partial
t}\,=\,v_\epsilon\,\frac{\partial\eta}{\partial\xi}\,+\,\frac{\partial^2\eta}{\partial\xi^2}\,+\,\frac{\delta
f(\phi)}{\delta\phi}\bigg|_{\phi\,=\,\phi_\epsilon}\,\eta\,.
\label{e6}
\end{eqnarray}
Since this equation is linear in $\eta$, the question of stability can
be answered by studying the spectrum of the temporal eigenvalues. To
this end, we express $\eta(\xi,t)$ as
\begin{eqnarray}
\eta(\xi,t)\,=\,e^{-\,Et}\,e^{-\,v_\epsilon\xi/2}\,\psi_E(\xi)\,,
\label{e7}
\end{eqnarray}
which converts Eq. (\ref{e6}) to the following one-dimensional
Schr\"odinger equation  for a particle in a  potential (with
$\hbar^2/2m=1$):
\begin{eqnarray}
\left[\,-\,\frac{d^2}{d\xi^2}\,+\,\frac{v^2_\epsilon}{4}\,-\,\frac{\delta
f(\phi)}{\delta\phi}\bigg|_{\phi\,=\,\phi_\epsilon}\right]\psi_E(\xi)\,=\,E\,\psi_E(\xi).
\label{e8}
\end{eqnarray}
In Eq. (\ref{e8}), the quantity
$\displaystyle{V(\xi)=\left[\,\frac{v^2_\epsilon}{4}\,-\,\frac{\delta
f(\phi)}{\delta\phi}\bigg|_{\phi\,=\,\phi_\epsilon}\right]}$ plays the
role of the potential. If we now denote by $\xi_0$ the coordinate of
the point where $\phi_\epsilon(\xi)=\epsilon$, then for the
nonlinearity (\ref{e2}) the potential $V(\xi)$ is easily seen to have
the form
\begin{eqnarray}
V(\xi)\,=\,\left[\,\frac{v^2_\epsilon}{4}-1+n\phi^{n\,-\,1}_\epsilon(\xi)\right]\Theta(\xi_0-\xi)\,-\frac{1}{v_\epsilon}\,\delta(\xi-\xi_0)\nonumber\\&&\hspace{-4cm}+\,\frac{v^2_\epsilon}{4}\,\Theta(\xi\,-\,\xi_0)\,.
\label{e10}
\end{eqnarray}
The $\delta$-function in Eq. (\ref{e10}) appears from the functional
derivative in Eq. (\ref{e6}), since there is a discontinuity of
magnitude $\epsilon$ in $f(\phi)$ at $\phi=\epsilon$. This
discontinuity contributes an amount equal to
\begin{equation}
\phi_\epsilon\,\frac{d \Theta
(\phi_\epsilon-\epsilon)}{d\phi_\epsilon}\,=\,\phi_\epsilon\,\delta(\phi_\epsilon-\epsilon)\,=\,\frac{\epsilon}{|\phi'_{\epsilon}(\xi_0)|}\,\delta
(\xi-\xi_0)
\end{equation}
to $V(\xi)$. If we combine this with the fact that
$|\phi'_{\epsilon}(\xi_0)|=\epsilon v_\epsilon$, which follows
immediately from the fact that one simply has  $\phi_{\epsilon}(\xi) =
\epsilon e^{-v_\epsilon (\xi-\xi_0)}$ for $\xi \ge \xi_0$, one obtains
the $\delta$-function term in the potential given in Eq. (\ref{e10}).

\begin{figure}
\begin{center}
\includegraphics[width=0.45\textwidth]{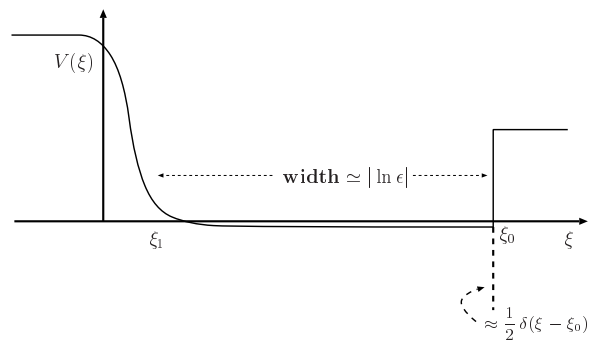}
\end{center}
\caption{The potential $V(\xi)$ in the Schr\"odinger operator obtained
in the stability analysis. \label{fig1}}
\end{figure}

The form of the potential $V(\xi)$ is sketched in
Fig. \ref{fig1}. Notice that $\phi_\epsilon(\xi)$ is a monotonically
increasing function from $\epsilon$ at $\xi_0$ {\it towards the  
left\/}, asymptotically reaching the value $1$ as
$\xi\rightarrow-\infty$. As a result, for $\xi<\xi_0$, $V(\xi)$ also
increases monotonically towards the left, from $v^2_{\epsilon}/4
-1+n\epsilon^{n-1}\simeq -\pi^2/\ln^2\epsilon$ at $\xi=\xi_{0-}$, to
$(n-\pi^2/\ln^2\epsilon)\approx n$ as $\xi\rightarrow-\infty$. At
$\xi_0$, there is an attractive $\delta$-function potential of
strength $1/v_{\epsilon}\approx 1/2$ and a finite step of height
1. The crucial feature for the stability analysis below is the fact
that $V(\xi)$ stays remarkably flat at a value $-\pi^2/\ln^2\epsilon$
over a distance $(\xi_0-\xi_1)\simeq|\ln\epsilon|$, and then on the
left of $\xi_1$, it increases to the value $\approx n$, over a
distance of order unity. As argued in Sec. II.B, this is a consequence
of the nature of the solution $\phi_\epsilon(\xi)$.

If there are negative eigenvalues of the above Schr\"odinger equation,
then according to Eq. (\ref{e7}) $\eta(\xi,t)$  grows in time in the
comoving frame, i.e.,  the front solution $\phi_\epsilon(\xi)$ is
unstable. On the other hand, if there are no negative eigenvalues,
then the asymptotic front shape is stable, and the spectrum of the
eigenvalues then determines the nature of the relaxation of
$\phi(x,t)$ to the solution $\phi_\epsilon(\xi)$.

The full spectrum in general depends on the boundary conditions
imposed on the eigenfunctions $\psi_E$. Here we consider only
localized perturbations, for which we need to have $\eta(\xi,t)\to 0$
as $\xi \to \pm \infty$. Because of the exponential factor in
Eq. (\ref{e7}), any eigenfunction $\psi_E$ which vanishes as $\xi \to
\infty$ is consistent with vanishing $\eta$ towards the
right\cite{notehilbert}. However, for $\xi\to -\infty$, the
eigenfunctions $\psi_E$ need to vanish exponentially fast with a
sufficiently large exponent, so that when it is combined with the
exponentially diverging term $e^{-v_\epsilon/2}$, they are still
consistent with the requirement that $\eta $ vanishes for
$\xi\to-\infty$. For the lowest ``energy'' eigenvalues, which we will
investigate below, we will demonstrate that these requirements are
obeyed.

\subsection{Shape of $\phi_\epsilon(\xi)$ and the Zero Mode
of the Stability Operator}

From the form in the potential, it is clear that the lowest ``energy''
eigenmodes, i.e., the slowest relaxation eigenmodes, are the ones that
are confined to the bottom of the potential. This is the region where
the nonlinear terms proportional to $\phi^{n-1}$ are negligible, and 
which is often called the ``leading edge'' of the front
profile. 
 For $\epsilon\ll1$, the solution of $\phi_\epsilon(\xi)$
in this leading edge  is given by \cite{bd}
\begin{eqnarray}
\phi_\epsilon(\xi)\,\approx\,\frac{|\ln\epsilon|}{\pi}\,\sin[z_i\xi]\,e^{-\,z_r\xi}\quad\!\!\mbox{for}\,\,\xi_1\lesssim\xi\leq\xi_0\simeq
|\ln\epsilon|\nonumber\\&&\hspace{-7.1cm}=\,\epsilon\,e^{-\,v_\epsilon(\xi\,-\,\xi_0)}\quad\quad\quad\quad\!\!\mbox{for}\,\,\xi\geq\xi_0\,.
\label{e9}
\end{eqnarray}
Here, $z_i\approx\pi/|\ln\epsilon|$ and $z_r=1+O(\epsilon^2)$. The
values of $\phi_\epsilon(\xi)$ and
$\displaystyle{\frac{d\phi_\epsilon}{d\xi}}$ are continuous at
$\xi=\xi_0$, and $\phi_\epsilon(\xi_0)=\epsilon$. Although
Eqs. (\ref{e9}) and (\ref{zeromode}) suggest at first sight that the
$\phi_{\epsilon}(\xi)$ has a node at $\xi=0$, Eq. (\ref{e9}) is only
valid in the leading edge, and $\phi_{\epsilon}(\xi)$ crosses over to
other behavior around $\xi_1$, which makes the front solution
$\phi_\epsilon(\xi)$ a mononically decreasing function of $\xi$. The
value of $\xi_1$ is set by the criterion that around $\xi_1$, the
nonlinear terms of $f(\phi_\epsilon(\xi))$ starts to become
significant, just like $\xi_1$ marks the point where the potential
$V(\xi)$ crosses over from the asymptotic value on the left to the
bottom value. The coordinate $\xi_1$, therefore, is more or less
fixed; on the other hand, $\xi_0$ asymptotically diverges as
$\simeq|\ln\epsilon|$ for small $\epsilon$, making $(\xi_0-\xi_1)$
also diverge as $\simeq | \ln\epsilon |$. This is an immediate
consequence of the overall exponential decay of $\phi_{\epsilon}(\xi)$
in $\xi$ at the leading edge.

From the form in the potential, it is clear that the lowest ``energy''
eigenmodes, i.e., the slowest relaxation eigenmodes, are the ones that
are confined to the bottom of the potential. We notice that among
these modes, invariably there is a zero mode of the stability operator
that is associated with the uniformly translating front solution of a
dynamical equation like Eq. (\ref{e1}): since  $\phi_\epsilon(\xi)$
and $\phi_\epsilon(\xi+a)$ are solutions of Eq. (\ref{e4}) for any
arbitrary $a$, we find by expanding to first order in $a$ that
$\psi_0(\xi)=e^{v_\epsilon\xi/2}\displaystyle{\frac{d\phi_\epsilon}{d\xi}}$
is a solution of Eq. (\ref{e8}) with eigenvalue $E=0$. From the result
(\ref{e9}) for the asymptotic front solution, we then immediately
get {\em to dominant order}
\begin{equation}
\psi_0 \sim \sin{z_i \xi} , ~~~~~z_i \simeq \pi/| \ln \epsilon |
,~~~~~\xi_1\lesssim\xi\leq\xi_0. \label{zeromode}
\end{equation}
Furthermore, since $\phi_\epsilon(\xi)$ is a monotonically decreasing
function of $\xi$, the solution
$\psi_0(\xi)=e^{v_\epsilon\xi/2}\displaystyle{\frac{d\phi_\epsilon}{d\xi}}$
is nodeless. Since we know from elementary quantum mechanics that the
nodeless eigenfunction has the lowest eigenvalue, this implies that
all the {\it other\/} eigenvalues of Eq. (\ref{e8}) are positive,
i.e., the solution $\phi_\epsilon(\xi)$ is stable.

The spectrum of eigenvalues of Eq. (\ref{e8}) for $E>0$, therefore, is
going to determine the decay property of localized perturbations
$\eta(\xi,t)$ in time. We notice that for $E>v_\epsilon^2/4 \approx
1$, the value of the potential on the far right, the spectrum of
eigenvalues will be continuous. However, we are particularly
interested in the smallest eigenvalues $E_m>0$ for small $m$, since
these are  the eigenmodes that decay the slowest in time. These are
the eigenvalues associated with bound states in the potential well.

\subsection{Lowest Eigenmodes and Eigenvalues for $\epsilon \ll 1$}

As $\epsilon\rightarrow0$, the bottom well of the potential becomes
very wide: its width diverges as $|\ln \epsilon|$. As we know from
elementary quantum mechanics, the lowest ``energy'' eigenfunctions
then become essentially sine or cosine waves in the potential well
with small wave numbers $k$ and correspondingly small ``energy''
eigenvalues.

\begin{figure}
\begin{center}
\includegraphics[width=0.45\textwidth]{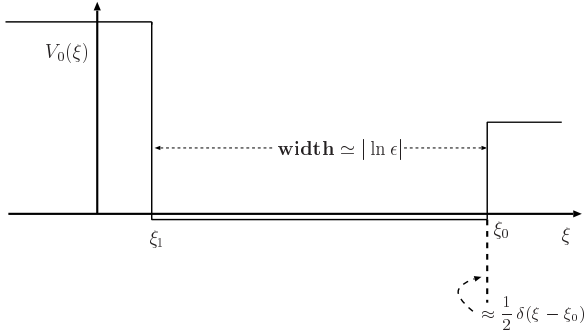}
\end{center}
\caption{The approximate potential $V_0(\xi)$ that can be used for
calculating the low-lying modes for large widths of the bottom well,
i.e., for small $\epsilon$. \label{fig2}}
\end{figure}

Based on the fact that the potential $V(\xi)$ on the left rises over
length scales of order unity, we now make an approximation. In the
limit that the bottom well is very wide and the  $k$ values of the
bound state eigenmodes very small, it becomes an increasingly good and
an asymptotically correct approximation to view the left wall of the
well simply as a steep step, as sketched in  Fig. \ref{fig2} --- we
thus approximate the potential by
\begin{eqnarray}
V_0(\xi)\,=\,n\,\left[\,1\,-\,\Theta(\xi)\right]\,-\,\frac{\pi^2}{\ln^2\epsilon}\,\Theta(\xi)\,\left[\,1\,-\,\Theta(\xi\,-\,\xi_0)\right]\,\nonumber\\&&\hspace{-5.3cm}-\,\frac{1}{2}\,\delta(\xi-\xi_0)\,+\,\Theta(\xi\,-\,\xi_0)\,.
\label{e11}
\end{eqnarray}

On the right hand side, there is an attractive delta-function
potential at the point where the potential shows a step to a value
close to 1. It is easy to check that the prefactor of the
delta-function of 1/2 is not strong enough to give rise to bound
states with $E<0$, and as a result, for very small values of
$\epsilon$, the low-lying eigenmodes  approach  sine waves with
nodes at the position of the walls of the potential \cite{note2},
\begin{equation}
\psi_m \simeq \sin\left[ k_m (\xi-\xi_1)\right].
\end{equation}

The condition that these solutions have nodes at the right edge of the
well then yields
\begin{equation}
k_m \simeq \frac{(m+1) \pi}{\xi_0-\xi_1} \simeq \frac{(m+1) \pi}{|\ln
\epsilon |},
\end{equation}
implying that the corresponding eigenvalues are given by
\begin{eqnarray}
E_m\,\simeq \,\frac{\left[
(m+1)^2\,-\,1\right]\,\pi^2}{\ln^2\epsilon}\,\quad\quad m\,=\,0,1,
2,\ldots
\label{e15}
\end{eqnarray}
Here, the first term between square brackets comes from the ``kinetic
energy'' term $k^2$, while the second term orgininates from  the value
of the potential at the bottom.

Note that for $m=0$, the eigenmode $\sin k_0$ with eigenvalue $E_0$ is
indeed the same as the zero eigenmode of Eq. (\ref{zeromode}) with
$k_0=z_i$, which we calculated from the shape of the front solution
$\phi_\epsilon$ in the leading edge. Besides verifying the consistency
of our approach, this also confirms that there are no corrections to
Eq. (\ref{e15}) for $m=0$: for $m=0$ it will yield an eigenvalue zero
to all orders in $\epsilon$. Therefore, the smallest nonzero
eigenvalue, which governs the relaxation of the front velocity and
profile to the asymptotic ones is $E_1$ with relaxation time $\tau_1$
given by
\begin{eqnarray}
\tau_1^{-1} = E_1 \simeq \,\frac{3 \,\pi^2}{\ln^2\epsilon}\,.
\label{e16}
\end{eqnarray}

Equation (\ref{e15}) also confirms that as $\epsilon\rightarrow0$, the
gap between the spectral lines decreases as $\ln^{-2}\epsilon$, which
is consistent with the fact that for a pulled front $\epsilon=0$ and
the spectrum becomes gapless. Also notice that for the eigenvalues in
Eq. (\ref{e15}), the corresponding eigenmodes $\psi_E(\xi)$ decay as
$\exp[-\sqrt{n}\,\,|\xi|]$ for $\xi\rightarrow-\infty$ and as
$\exp[-v_\epsilon\xi/2]$ for $\xi\rightarrow\infty$, which make
$e^{v_\epsilon\xi/2}\psi_E(\xi)$ to go to zero for
$\xi\rightarrow\pm\infty$, satisfying the boundary conditions
discussed  previously at the end of section II.A.

\ecols

\begin{thebibliography}{99}
\bibitem{notehilbert} The fact that eigenfunctions $\psi_E$ which
diverge as $\xi\to \infty$ are allowed, means that there are
admissable eigenfunctions which are not in the Hilbert space of the
Schr\"odinger operator. See e.g., Ref. \onlinecite{ebert} for further
discussion of this point.
\bibitem{note2} More explicitly, if we write the solution within the
well as $A\sin [k (\xi-\xi_1) +B]$, and for $\xi > \xi_0$ as $A_+
e^{-(\xi-\xi_0)}$ (which is correct to lowest order in $\epsilon$),
then we get from the boundary conditions at $\xi_0$: $2k\,
\mbox{cotg} [k(\xi_0-\xi_1)+B] = -1$; likewise, at the left boundary,
of the well, we get $k \, \mbox{cotg}\,B = const.$, where the constant is
determined by the size of potential step. For $\xi_0-\xi_1\gg1$, the
small-$k$ solutions are those stated in the text with $B \to 0$.
\end{thebibliography}
\end{document}